\documentclass[aps,prd,superscriptaddress,reprint,floatfix,nofootinbib,preprintnumbers]{revtex4-1}
\usepackage{dcolumn}   
\usepackage{epsfig}
\usepackage{rotating}
\usepackage[usenames, dvipsnames]{color}
\usepackage{mathrsfs}
\usepackage{tipa}
\usepackage{bm}
\usepackage{bbm}
\usepackage{amsmath}
\usepackage{amssymb}
\usepackage{amsthm}
\usepackage{amsfonts}
\usepackage{mathtools}
\usepackage{latexsym}
\usepackage{relsize}
\usepackage[greek,english]{babel}
\usepackage{booktabs}
\usepackage[iso-8859-7]{inputenc}

\usepackage{array}
\newcolumntype{?}{!{\vrule width 2pt}}

\newcommand{\lp}{\left(}
\newcommand{\rp}{\right)}

\newcommand{\lsim}   {\mathrel{\mathop{\kern 0pt \rlap
  {\raise.2ex\hbox{$<$}}}
  \lower.9ex\hbox{\kern-.190em $\sim$}}}
\newcommand{\gsim}   {\mathrel{\mathop{\kern 0pt \rlap
  {\raise.2ex\hbox{$>$}}}
  \lower.9ex\hbox{\kern-.190em $\sim$}}}
\newcommand{\bw}{\begin{widetext}\begin{equation}}
\newcommand{\ew}{\end{equation}\end{widetext}}
\newcommand{\be}{\begin{equation}}
\newcommand{\ee}{\end{equation}}
\newcommand{\ba}{\begin{eqnarray}}
\newcommand{\ea}{\end{eqnarray}}

\newcommand{\nn}{\nonumber}

\begin{document}

	\title{The spectrum of symmetric teleparallel gravity}
	\author{Aindri\'u Conroy}
	\affiliation{Physics Department, Lancaster  University, Lancaster, LA1 4YB, UK}
	\author{Tomi Koivisto}
	\affiliation{Nordita, KTH Royal Institute of Technology and Stockholm University, Roslagstullsbacken 23, 10691 Stockholm, Sweden}
	\date{\today}
	\preprint{NORDITA 2017-101}
	
	\begin{abstract}

		General Relativity and its higher derivative extensions have {\it symmetric teleparallel} reformulations in terms of the non-metricity tensor within a torsion-free and flat 				geometry. These notes present a derivation of the exact propagator for the most general infinite-derivative, even-parity and generally covariant theory in the symmetric 				teleparallel spacetime.  The action made up of the non-metricity tensor and its contractions is decomposed into terms involving the metric and a gauge vector field and is found 		to complement the previously known non-local ghost- and singularity-free theories.

	\end{abstract}
	
	\maketitle
	
	Recent detections of gravitational waves \cite{Abbott:2016blz, TheLIGOScientific:2017qsa}, or fluctuations in the gravitational field, fully agree with the predictions of General Relativity (GR). As a theory of the metric gravitational field, however, GR remains incomplete in the ultra-violet. 
		Simple but infinite-derivative-order actions that alleviate the singular structure of 	
		GR without introducing new degrees of freedom \cite{Krasnikov:1987yj,Tomboulis:1997gg,Modesto:2011kw,Biswas:2011ar}, have lead to promising results in recent investigations 
		into e.g. quantum loops \cite{Talaganis:2014ida,Modesto:2015lna}, scattering amplitudes \cite{Dona:2015tra,Talaganis:2016ovm}, inflation \cite{Briscese:2013lna,Koshelev:2016xqb}, bouncing cosmology 	
		\cite{Biswas:2005qr,Biswas:2010zk, Conroy:2016sac} and black holes \cite{Frolov:2015bta, Conroy:2015wfa}.

	The purpose of this note is to generalise the classification of metric theories from Riemannian (see Ref. \cite{Biswas:2011ar}) to a more general geometry.
	Recently it has been suggested that a reconciliation between gravitation as a gauge theory of translations \cite{Feynman:1996kb,deAndrade:2000kr} and as a gauge theory of the general linear transformation $GL(4)$ \cite{1971AnPhy..62...98I,Hehl:1994ue} could be achieved by stipulating that the former group of transformations should be the unbroken remainder of the latter, in the frame where inertial effects are absent \cite{us} (where by `unbroken' we mean that the gauge is fixed in such a way that the affine connection always remains a translation). 
	This logic leads us to the symmetric teleparallel spacetime \cite{Nester:1998mp}, see also \cite{Adak:2005cd,Adak:2008gd,Mol:2014ooa,us}.

\begin{figure}[h]
		\includegraphics[scale=0.35]{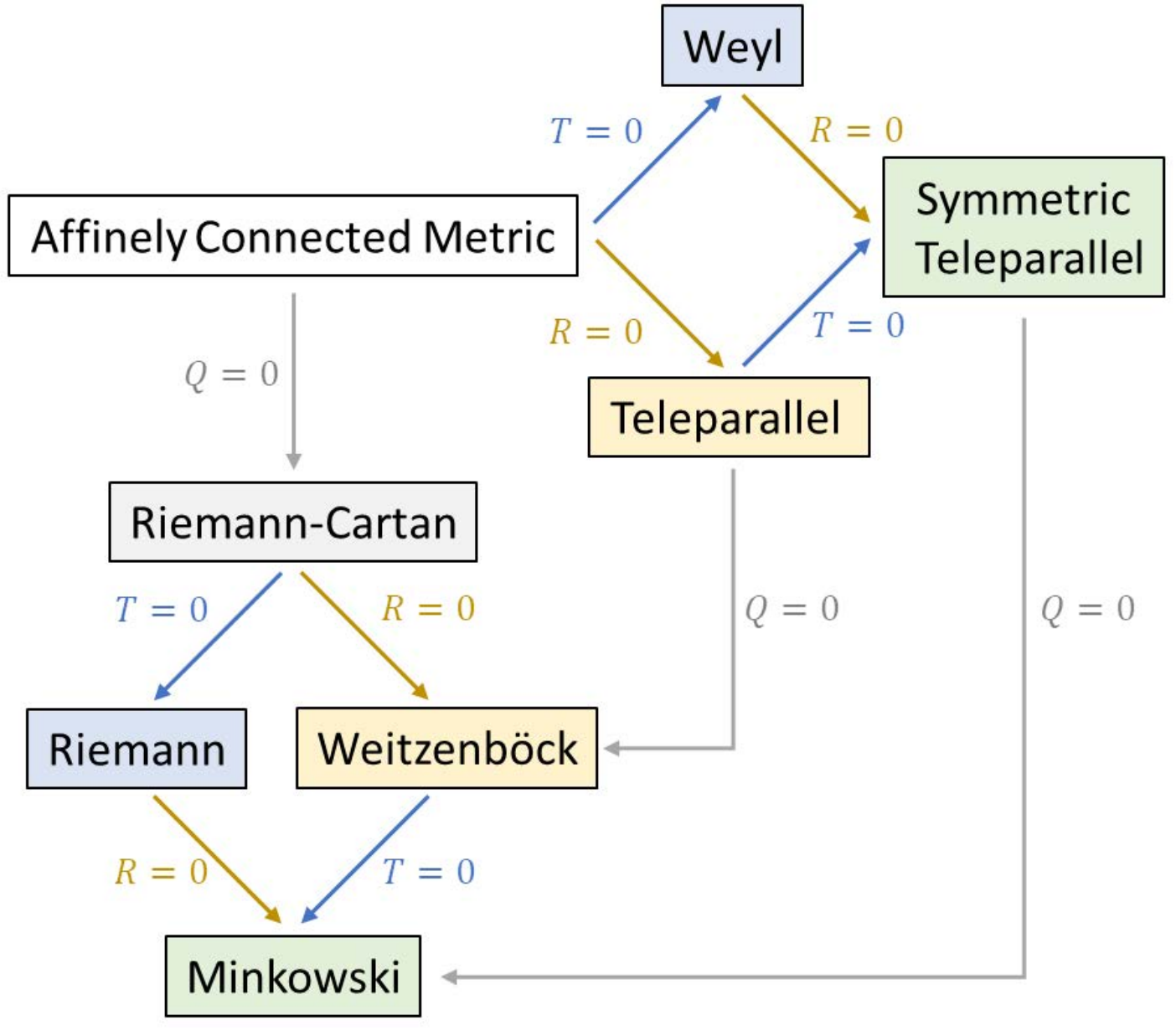}	
		\caption{Diagram of affinely connected metric spacetimes illustrating all permutations of the non-metricity tensor $Q_{abc}$, the Riemann curvature $R^a{}_{bcd}$ and the torsion $T^a{}_{bc}$, where indices have been suppressed for presentation purposes. For more details on these spacetimes see Refs. \cite{deAndrade:2000kr,1971AnPhy..62...98I,Hehl:1994ue,us,Nester:1998mp,Adak:2005cd,Adak:2008gd,Mol:2014ooa,Kibble:1961ba,Tresguerres:2000qn,Golovnev:2017dox}.}
		\label{fig:figspacet}
	\end{figure}

	An affine connection $\Gamma^a_{\phantom{a}bc}$ is invariantly characterised by its curvature,
	\be \label{riemann}
	{R}^a_{\phantom{a}bcd} = 
	2\partial_{[c} \Gamma^a_{\phantom{a}d]b}
	+ 2\Gamma^a_{\phantom{a}[c\lvert e\rvert}\Gamma^e_{\phantom{e}d]b}\,,
	\ee
	and its torsion, $T^a_{\phantom{a}bc}=2\Gamma^a_{\phantom{a}[bc]}$. In a \emph{teleparallel} spacetime, where 
	$R^a{}_{bcd}=0$, the inertial connection is given by a general linear transformation $J^{a}{}_{b}$ of the trivial vanishing connection solution or ``coincident gauge'' \cite{us}, as 
	$\Gamma_{\phantom{a}bc}^{a}=(J^{-1})^{a}{}_{d}\partial_{b}J^{d}{}_{c}$, where $(J^{-1})^{a}{}_{d}$ are the components
	of the inverse matrix that parameterises the $GL(4)$ transformation. In a \emph{symmetric teleparallel} spacetime, the torsion also vanishes $T^a_{\phantom{a}bc}=0$. It follows that $(J^{-1})^{a}{}_{d}\partial_{[b}J^{d}{}_{c]}=0$, which indeed 
	leaves us with the coordinate-changing diffeomorphism $J^{a}{}_{b} =\partial_b \xi^a$ which was identified with translations (by a vector $\xi^a$ in the tangent space) in the gauging of the Poincar\'e group already in Ref. \cite{Kibble:1961ba}. The frame field can be obtained by the nonlinear realisation of the translation gauge potential \cite{Tresguerres:2000qn}. We refer the reader to Fig. \ref{fig:figspacet} for the relations between the eight types of affinely connected spacetimes. 
	
	In the presence of a metric $g_{ab}$, we can not only define the curvature and torsion but also the non-metricity $Q_{abc}=\nabla_a g_{bc}$, where $\nabla_a$ is the covariant 		derivative with respect to the
	affine connection $\Gamma^a_{\phantom{a}bc}$. 
	The non-metricity has two independent traces, which we denote as 
	$Q_a=Q_{a\phantom{a}b}^{\phantom{a}b}$ and $\tilde{Q}_a=Q_{b\phantom{b}a}^{\phantom{b}b}$. The quadratic form 
	\be \label{q2}
	Q^2 =   \frac{1}{4}Q_{abc}Q^{abc} -  \frac{1}{2}Q_{abc}Q^{bca} 
	-   \frac{1}{4}Q_a Q^a  
	+ \frac{1}{2}Q_a\tilde{Q}^a\,,
	\ee
	is equivalent to the metric Ricci scalar $\mathcal{R}$, up to a boundary term \cite{us}. To be more precise, $\mathcal{R}$ is contraction of the Riemann tensor (\ref{riemann}) via $\mathcal{R}= g^{ac}g^{bd}\mathcal{R}_{abcd}$, where the affine connection $\Gamma^a{ }_{bc}$ is none other than the Christoffel symbol, while the boundary term is given by \cite{us} $\mathcal{D}_a(Q^a-\tilde{Q}^a)$, where $\mathcal{D}_a$ is the covariant derivative
	of the Christoffel symbol. 
	
	Beltr\'an {\it et al} \cite{us} introduced the Palatini formalism for teleparallel and symmetric teleparallel gravity theories. 
	In these notes, however, we adopt the inertial variation
	\cite{Golovnev:2017dox} as the recipe to obtain the field equations in a covariant form within the desired geometry. 
	Our degrees of freedom are the fluctuations $h_{ab}$ in the metric $g_{ab}$, and the vector fluctuations $u^a$ that determine the connection $\Gamma^c{}_{ab}$,
	\be \label{expansion}
	g_{ab}=\eta_{ab}+h_{ab}\,, \quad \Gamma_{cab}=\partial_{a}\partial_{b}u_{c}\,.
	\ee
	From the above, it is straightforward to verify that the components of the non-metricity tensor are given by 
	\begin{equation}
	\label{Qexpansion}
	Q_{abc}=\partial_{a}(h_{bc}-2\partial_{(b}u_{c)}),
	\end{equation}
	which is by construction a tensor and invariant under the infinitesimal diffeomorphism 
	$\delta_\xi h_{ab} =2\xi_{(a,b)} $, $\delta_\xi u_a = \xi_a$. 

	We consider the most general covariant, parity-even Lagrangian that is quadratic in non-metricity\footnote{For the purpose of deriving the propagator for an arbitrary theory, it suffices to expand the action to the quadratic order in the field strength. Furthermore, the action is parity-even if it is invariant under parity-transformations whereby the sign of the spatial coordinates is flipped.}. There are 10 possible terms: 
	\ba 
	-2{\cal L}_G & = & -\frac{1}{2}\left[Q^{abc}a_{1}(\Box)+2Q^{bac}b_{1}(\Box)\right]Q_{abc} \nn \\
	& + & \left[\nabla_{e}Q^{aec}b_{2}(\Box)+\frac{1}{2}\nabla_{e}Q^{cea}f_{2}(\Box)\right]\Box^{-1}\nabla^{b}Q_{abc}\nonumber \\
	& - & \left[Q^{a}c_{2}(\Box)+\frac{1}{2}\tilde{Q}^{a}f_{3}(\Box)\right]\Box^{-1}\nabla^{b}\nabla^{c}Q_{abc} \nn \\
	& - & \frac{1}{2}\left[Q^{a}d_{1}(\Box)Q_{a}+2\tilde{Q}^{a}c_{1}(\Box)Q_{a}+2\tilde{Q}^{a}b_{3}(\Box)\tilde{Q}_{a}\right]\nonumber \\
	& - & \frac{1}{2}Q^{aef}f_{1}(\Box)\Box^{-2}\nabla_{e}\nabla_{f}\nabla^{b}\nabla^{c}Q_{abc}\,. \label{lagrangian}
	\ea
	The functions $a_i(\Box)$, $b_i(\Box)$, $c_i(\Box)$, $d_i(\Box)$, $f_i(\Box)$ are analytic functions of the D'Alembertian operator $\Box\equiv g^{ab}\nabla_a \nabla_b$, modulated 		by a mass scale $M$, so as to remain dimensionless. The special case with the five non-vanishing constants $a_1$, $b_1$, $c_1$, $d_1$ and $b_3$ has been considered 		previously \cite{Adak:2005cd,Adak:2008gd,us}.

	We include matter sources by considering the total Lagrangian density to be given by $\mathcal{L} = \mathcal{L}_G + \mathcal{L}_M$\,,
	while taking into account the currents at the linear order in the perturbative expansion \eqref{expansion}, like so
	\be \label{hyper}
	\tau_{ab} = \frac{-2}{\sqrt{-g}} \frac{\delta{\lp \sqrt{-g}\mathcal{L}_M\rp}}{\delta g^{ab}}\,, \quad \tau_{a} =  \frac{-2}{\sqrt{-g}}\frac{\delta{\lp \sqrt{-g}\mathcal{L}_M\rp}}{\delta u^a}\,.
	\ee
	These are the (linearised) stress energy tensor and the hyperstress vector, respectively. 
	By substituting the expansion (\ref{Qexpansion}) into the Lagrangian (\ref{lagrangian}) and varying w.r.t. $h_{ab}$ and $u_{a}$, we
	obtain the equations of motion for the two fields.
	Thus, the field equation for the metric field can be written as
	\ba
	-\tau^{ab} & = & a\Box\left(h^{ab}-2\partial^{(a}u^{b)}\right) \nn \\ & + & 2b\left(\partial^{c}\partial^{(a}h_{c}{}^{b)}-\partial^{(a}\Box u^{b)}-\partial^{c}\partial^{a}\partial^{b}u_{c}\right)\nonumber \\
	& + & c\left[\partial^{a}\partial^{b}h-2\partial^{a}\partial^{b}\partial^{d}u_{d}+\eta^{ab}\left(\partial^{e}\partial^{c}h_{ec}-2\partial^{c}\Box u_{c}\right)\right]\nonumber \\
	& + & \eta^{ab}d\Box\left(h-2\partial^{c}u_{c}\right) \nn \\
	& + & f\Box^{-1}\left(\partial^{a}\partial^{c}\partial^{e}\partial^{b}h_{ec}-2\partial^{a}\partial^{b}\partial^{c}\Box u_{c}\right)\,, \label{eq:metriceoms}
	\ea
	while the field equation for the connection field is given by
	\ba
	-\frac{1}{2}\tau^{a} & = & \left[a+b\right]\Box\left(\partial^{b}h_{b}{}^{a}-\Box u^{a}\right) \nn \\
	& - & \left[(a+b)+2(c+d)+2(b+c+f)\right]\Box\partial^{c}\partial^{a}u_{c}\nonumber \\
	& + & \left[b+c+f\right]\partial^{b}\partial^{c}\partial^{a}h_{bc}+\left[c+d\right]\Box\partial^{a}h\,. \label{taueom}
	\ea
	 The reader should note that we have eased the burden of notation by dropping the arguments from the 10 functions appearing in (\ref{lagrangian}) and by grouping them
	into the 5 functions defined by
	\ba
	a & \equiv & a_{1}\,,\nonumber \\
	b & \equiv & b_{1}+b_{2}+b_{3}\,,\nonumber \\
	c & \equiv & c_{1}+c_{2}\,,\nonumber \\
	d & \equiv & d_{1}\,,\nonumber \\
	f & \equiv & f_{1}+f_{2}+f_{3}\,. \label{letters}
	\ea 
	
	The first port of call is to establish consistency with known results in the familiar Riemannian geometry. To this end, let us turn our attention to what happens if we eliminate the vector field $u^a$ from the theory. Substituting $u^a=0$ into (\ref{eq:metriceoms}) and we find, as expected, that the metric field equation does indeed reduce to the known field equation in Riemannian geometry, see \cite{Biswas:2011ar}. Furthermore, it is straightforward to check that the Bianchi identity i.e. $\partial_a \tau^{ab}=0$ dictates that the constraints 
	\be \label{diff}
	u^a =0\quad \Rightarrow\quad a+b=c+d=b+c+f=0\,,
	\ee
	which were established in \cite{Biswas:2011ar} hold also in this case. Moreover, substitution of these constraints into (\ref{taueom}) results in the field equations vanishing identically in the absence of a source, i.e. $\tau^{a}=0$. 
	
	Having obtained commonality with known results for the most general, parity-even action that is quadratic in curvature within Riemannian geometry, we now turn our attention to GR. From (\ref{q2}), we know that the equivalent of the Einstein-Hilbert Lagrangian is given by $\mathcal{L}_{G} = Q^2$. Upon reflection, we observe that the Lagrangian (\ref{lagrangian}) reduces to that of GR for the non-vanishing parameters $a_1=-b_1=c_1=-d_1=1$, which obey (\ref{diff}), as required. 
	%
	At the linear order, there are indeed many other equivalent theories that remain
	invariant consistently within the $\Gamma^a{}_{bc}=0$ gauge, i.e. (\ref{diff}) is satisfied. To be explicit, the action with five free functions, 
	\ba \label{greq}
	{\cal L}_{G}  & = &  Q^2 + b_2(\Box)\lp \tilde{Q}^aQ_a -2 Q^{bac}Q_{abc}\rp  \nn \\ 
	& + & b_3(\Box)\lp \tilde{Q}^aQ_a - \nabla_e Q^{aec}\Box^{-1}\nabla^b Q_{abc}\rp \nn \\
	& + & c_{1}(\Box){Q}^{a}\lp \tilde{Q}_{a} - \Box^{-1}\nabla^{b}\nabla^{c}Q_{abc}\rp\nn \\
	& + & f_2(\Box)\lp \nabla_e Q^{cea} + \tilde{Q}^a\nabla^{b}\rp\Box^{-1}\nabla^{c}Q_{abc} \nn \\
	& + & f_{3}(\Box)\lp \tilde{Q}^a - Q^{aef}\frac{\nabla_{e}\nabla_{f}}{\Box}\rp \frac{\nabla^{b}\nabla^{c}}{\Box}Q_{abc}  \,, 
	\ea 
	represents the class of theories that reduce to (symmetric teleparallel) equivalents of GR at the quadratic order. 
	
	Concerning higher derivatives, we note that the connection equation of motion\footnote{It can be
	seen from Eq.(5.8.21) of Ref. \cite{Hehl:1994ue} that in the frame formulation, the case with non-metricity and vanishing torsion leads to higher-derivatives in the first (i.e. the frame) field
	equation, whereas in our Palatini formulation they appear in the second (i.e. connection) field equation.} is generically third order in metric derivatives unless parameters are chosen in such a way that (\ref{diff}) 	holds. Investigating whether this would pose a problem for the Cauchy formulation is beyond the scope of this study. 
	

	
	
	By taking into account the inertial connection, we retain consistency with the Bianchi identities for any choice of parameters to all orders \cite{us}.
	We demonstrate this important fact, at the linear order, by taking the divergence of (\ref{eq:metriceoms}),
	\ba
	-\partial_{a}\tau^{ab} & = & (a+b)\Box\left(\partial_{a}h^{ab}-\Box u^{b}-\partial^{c}\partial^{b}u_{c}\right) \nn \\ & + & (c+d)\left(\Box\partial^{b}h-2\Box\partial^{c}\partial^{b}u_{c}\right)\nonumber \\
	& + & (b+c+f)\left(\partial^{c}\partial^{e}\partial^{b}h_{ec}-2\Box\partial^{c}\partial^{b}u_{c}\right)\,.\label{eq:divtab}
	\ea
	Comparison with (\ref{taueom}) confirms the relation
	\be \label{div}
	\partial_{a}\tau^{ab}=\frac{1}{2}\tau^{b}\,,
	\ee
	which is our desired conservation law. 
	At this point, we make a further simplification by assuming that matter is \emph{minimally coupled} in that the field $u^a$ does not enter into the 
	matter lagrangian $\mathcal{L}_M$. This is identically true in a vacuum and for canonical scalar and vector fields. To couple spinor matter one may assume the equivalent of the standard Levi-Civita connection recast in non-metric geometry \cite{Adak:2008gd}. The standard metric coupling is a consequence of the Hermitian property of the Dirac action \cite{Koivisto:2018aip}.
	
	Our intention now is to express the field equation \eqref{eq:metriceoms} purely in terms of the metric $h_{ab}$. In the minimally coupled system, we can neglect the hyperstress vector (\ref{hyper}) (i.e. $\tau^a=0$) and integrate out the connection.  
	By the divergence of the equation of motion (\ref{taueom}), we can relate the divergence of the vector $u^a$ to the derivatives of the metric fluctuation like so, 
	\be
	\Box\partial^{c}u_{c}  =  \frac{\alpha-(c+d)}{2\alpha}\partial_{e}\partial^{d}h_{d}{}^{e}
	+  \frac{\left(c+d\right)}{2\alpha}\Box h\,. \label{eq:hurel1}
	\ee
	Further still, by returning this result into the equation of motion (\ref{taueom}), it is possible to express the vector $u^a$ purely in
	terms of the derivatives of the metric fluctuation,
	\ba
	& - & \left(a+b\right)\Box^{2}u^{a}  =  -\left(a+b\right)\Box\partial^{b}h_{b}{}^{a}
	- \Big\{(b+c+f)  \nonumber \\ & - &  \frac{\left[2\alpha-(a+b)\right]\left[\alpha-(c+d)\right]}{2\alpha}\Big\}\partial^{b}\partial^{c}\partial^{a}h_{bc} \nn \\
	& - &  \Big\{(c+d) -  \frac{\left[2\alpha-(a+b)\right]\left[c+d\right]}{2\alpha}\Big\}\partial^{a}\Box h\,. \label{eq:hurel2}
	\ea
	Here, we have defined a short-hand notation introducing a parameter $\alpha$ that vanishes in the pure-metric case (\ref{diff}), 
	\be
	\alpha \equiv (a+b)+(c+d)+(b+c+f)\,.
	\ee
	As a cross-check of these relations, we note that (\ref{eq:hurel1}) and (\ref{eq:hurel2}) do indeed result in a vanishing divergence of the field equation (\ref{eq:metriceoms}), in accordance with the condition (\ref{div}) in our minimally coupled prescription $\tau^a=0$. To eliminate the inertial connection from (\ref{eq:metriceoms}) entirely, we require the
	combination $\partial^{(a}\Box u^{b)}$ which can be easily deduced from (\ref{eq:hurel2}).

	Substitution then reveals the field equations with the source $\tau^{ab}$ purely in terms of the field $h^{ab}$:
	\ba
	-\tau^{ab} & = & a\left(\Box h^{ab}-2\partial^{(a}\partial^{c}h_{c}{}^{b)} + \Box^{-1}\partial^{a}\partial^{c}\partial^{e}\partial^{b}h_{ec}\right)\nonumber \\
	& + & \left[d-\frac{(c+d)^{2}}{\alpha}\right]\Big[\eta^{ab}(\Box h-\partial^{e}\partial^{c}h_{ec})
	\nn \\ & - &  \partial^{a}\partial^{b}h  +  \Box^{-1}\partial^{a}\partial^{c}\partial^{e}\partial^{b}h_{ec}\Big]\,. \label{eq:acform}
	\ea
	As a final piece of book-keeping, let us define 
	\ba
	A & \equiv & a\,, \nn \\ 
	C & \equiv & \frac{(c+d)^{2}}{(a+b)+(c+d)+(b+c+f)} -d\,.  \label{params}
	\ea
	so that the field equation (\ref{eq:acform}) can be expressed in the following useful form 
	\ba \label{feom}
	-\tau^{ab} & = & A(\Box)\left(\Box h^{ab}-2\partial^{(a}\partial^{c}h_{c}{}^{b)}\right) \nn \\ & - & C(\Box)\left(\eta^{ab}\Box h-\eta^{ab}\partial^{e}\partial^{c}h_{ec}-\partial^{a}\partial^{b}h\right)
	\nn \\ & + & \left[A(\Box)-C(\Box)\right]\Box^{-1}\partial^{a}\partial^{c}\partial^{e}\partial^{b}h_{ec}\,.
	\ea
	We now move on to our main task, the propagator for the theory (\ref{lagrangian}). The metric propagator $\Pi_{abcd}$ is defined via $\Pi^{-1}_{abcd}h^{cd} = \tau_{ab}$ and is determined by the above form of the field equation
	\cite{Biswas:2011ar} in terms of the Barnes-Rivers spin projectors \cite{VanNieuwenhuizen:1973fi}. 
	If we first define the transversal projectors for the wavevectors $k^a$ in the Fourier space $\Box \rightarrow -k^2$ as
	\be
	\theta_{ab} = \eta_{ab} -  \frac{k_a k_b}{k^2}\,,
	\ee
	we can present the two projectors relevant to us as
	\be
	P^{(2)}_{abcd} =   \theta_{c(a}\theta_{b)d} - \frac{1}{2}\theta_{ab}\theta_{cd}\,, \quad
	P^{(0)}_{abcd}  =   \frac{1}{3}\theta_{ab}\theta_{cd}\,, \label{spin0}
	\ee
	for spin-2 and spin-0, respectively.
	The propagator, which is well-known to allow no vector excitations in flat space, is given by
	\be
	\Pi_{abcd}=  \frac{P^{(2)}_{abcd}}{A k^2} +\frac{3\lp A- C\rp}{2A\lp A - 3C \rp} \frac{P^{(0)}_{abcd}}{k^2} \,. \label{propagator}
	\ee
	Depending on the functions $A(\Box)$ and $C(\Box)$ in (\ref{params}), the spectrum can contain an arbitrary number of spin-2 and spin-0 modes.
	The GR form (\ref{q2}) is given by $A(-k^2)=C(-k^2)=1$, for which the scalar sector decouples. It is then the $k^{-2}{P^{(2)}_{abcd}}$ sector that propagates the graviton as in the 		massless representation with two polarisations. 

We have now isolated the propagator for the metric field but this is not the full story. We arrived at the result (\ref{propagator}) by integrating out the connection, using its own equation of motion (\ref{taueom}). In general, the reverse can not be similarly achieved, i.e. we cannot integrate out the metric to isolate the propagator for the connection due to the non-trivial coupling of the two fields. We can however arrive at the propagator for
the combined field content of the theory. 

This is most conveniently written in terms of $P^{(\bar{2})}$, the conventional projector of the spin-$2$ field inclusive of the longitudinal mode, \cite{VanNieuwenhuizen:1973fi}; $P^{(1)}$, which is the conventional vector projector; $P^{(\bar{0})}$ which projects the longitudinal scalar d.o.f. carried by the vector $u^a$ in addition to the time-like scalar d.o.f. projected by $P^{(0)}$; and $P^{(\times)}$ which describes the mixing between the two latter (we have removed subscripts $_{abcd}$ that are common to all the projectors). The full result for the total propagator is 
\ba \label{prop2}
k^2 \Pi & = &  \frac{1}{a}P^{(\bar{2})} + \frac{1}{a-b}P^{(1)} \\ &+& q_0^{-2}\Big[ \alpha P^{(0)} 
 +   (a+3d) P^{(\bar{0})} - \sqrt{3}(c+d) P^{(\times)}\Big]\,,\nonumber
\ea
where we have introduced the short-hand $q^2_0=(a+3d)\alpha-3(c+d)^2$. 

As expected, the field $u^a$ can carry a spin-$1$ d.o.f. It was explained and confirmed by various considerations in Refs. \cite{VanNieuwenhuizen:1973fi,Alvarez:2006uu} that pathologies cannot be avoided unless $a+b=0$ so that $P^{(1)}$-part of the propagator disappears.
As such, this is a constraint also upon the parameters of our theory. Furthermore, we require 
$a>0$ so as to avoid a ghost-like pole for the graviton. While the signs of the two remaining scalar pieces in the propagator (\ref{prop2}) should be chosen so as to avoid negative residues, we cannot exclude the possibility of these pieces remaining in a viable theory. 
To start with, the Newtonian limit \cite{Conroy:2014eja} 
	quite strictly constrains the additional scalar modes in the cases that do not satisfy (\ref{diff}). To categorically exclude such models, perhaps based on the possible difficulties with Cauchy formulation noted from  
	(\ref{taueom}), or by the  
	potential issues such as acausal propagation that may be activated at a nonlinear order, see e.g. \cite{Chen:1998ad,Deser:2014fta}, presents a considerable technical challenge. 
	
	It is clear that the equivalents of all Riemannian metric theories are contained within the symmetric teleparallel geometry, since the propagators of the latter are included in (\ref{propagator}). 	
	We note also that some of the potentially viable scalar-tensor theories contained in the action (\ref{lagrangian}) would, in the Riemannian context, require resorting to non-analytic functions with problematical integral operators such as $1/\Box$ \cite{Deser:2007jk,Conroy:2014eja}, which is due to the higher derivatives of the Riemannian invariants. 
		However, it is impossible to give a mass to the graviton when restricting to analytic functions, which is easy to see since a massive graviton would require a propagator of the form $P^{\bar{2}}/(k^2+m^2)$. Massive scalar fluctuations of the form $P^0/(k^2+m^2)$ are permitted, but the mass should be sufficiently large so as not to run into conflict with local constraints on long-range forces.

	Finally, we consider the case of infinite-derivative actions. By modulating the massless GR propagator by a suitable\footnote{For a suitable function: 1) $A(-k^2\rightarrow 0) \rightarrow 1$ so that GR is recovered in the infra-red, 2) $A(\Box)=e^{\gamma(\Box)}$, where $\gamma(\Box)$ is an entire function, so there are no additional poles, and 3) falls off sufficiently fast
		as $A(-k^2 \rightarrow -\infty) \rightarrow \infty$ so as to tame ultra-violet singularities. In the following examples, $\gamma({\Box})=-\Box/M^2$.} function, such as $A(\Box)=C(\Box)= e^{-\frac{\Box}{M^2}}$, we can improve the scaling $\sim k^{-2}$ of the GR propagator that leads to divergences in the ultra-violet. This is realised by the following infinite-derivative generalisation of the symmetric teleparallel equivalent:
\begin{eqnarray}
\label{qexp}
{\cal L}_G = \frac{1}{4} Q_{abc} e^{-\frac{\Box}{M^2}}P^{abc}\,, 
\end{eqnarray}
where 
\begin{equation}  
P_{abc} = \frac{1}{2}Q_{abc} - Q_{(cb)a} - \frac{1}{2}Q_a g_{bc} + \tilde{Q}_{(b} g_{c)a}\,.
\label{joku}
\end{equation}
Another possible theory, which at the linear order is equivalent to \eqref{qexp}, has an even simpler appearance, like so
	\be
	\label{qexp2}
	\mathcal{L}_G   =    \frac{1}{4}Q e^{-\frac{\Box}{M^2}} Q\,.
	\ee
	Both the above Lagrangians reproduce the metric dynamics of the Riemannian theory  \cite{Biswas:2011ar}
	\be \label{rexp}
	\mathcal{L}_G \sim \mathcal{R} + \mathcal{R}_{ab}\lp \frac{\exp{(-\frac{\mathcal{D}^2}{M^2})}-1}{\mathcal{D}^2}\rp \lp\mathcal{R}^{ab}-\frac{1}{2}\mathcal{R}\rp\,.
	\ee 
	Thus, the Lagrangians (\ref{qexp}), (\ref{qexp2}) and (\ref{rexp}) all have the same propagator in flat space: $\Pi_{abcd}= e^{\frac{-k^2}{M^2}}k^{-2}P^{(2)}_{abcd}$.

	The elegance of the new formulations gives rise to optimism for technical progress in the investigations into infinite-derivative gravity \cite{Modesto:2011kw,Biswas:2011ar,Talaganis:2014ida,Modesto:2015lna,Dona:2015tra,Talaganis:2016ovm,Briscese:2013lna,Koshelev:2016xqb,Biswas:2005qr,Biswas:2010zk,Conroy:2016sac,Frolov:2015bta,Conroy:2015wfa},
but also hint at a possible shortcut towards a finite quantum theory.
Non-locality has been recognised as a key to reconcile unitarity
with renormalisability \cite{Krasnikov:1987yj,Tomboulis:1997gg,Modesto:2011kw,Biswas:2011ar}. In the newly {\it purified gravity} \cite{us}, we
further avoid the conceptual difficulty of reconciling the
local character of the equivalence principle and the 
the non-local character of the quantum uncertainty principle \cite{Lammerzahl:1996se,Chiao:2003es}. In teleparallel gravity \cite{deAndrade:2000kr}, in contrast to GR, it is possible to separate the inertial effects from gravitation and to consider its quantisation. This separation is in-built to our geometry.

	\bibliography{Qletter}

\begin{thebibliography}{38}%
\makeatletter
\providecommand \@ifxundefined [1]{%
 \@ifx{#1\undefined}
}%
\providecommand \@ifnum [1]{%
 \ifnum #1\expandafter \@firstoftwo
 \else \expandafter \@secondoftwo
 \fi
}%
\providecommand \@ifx [1]{%
 \ifx #1\expandafter \@firstoftwo
 \else \expandafter \@secondoftwo
 \fi
}%
\providecommand \natexlab [1]{#1}%
\providecommand \enquote  [1]{``#1''}%
\providecommand \bibnamefont  [1]{#1}%
\providecommand \bibfnamefont [1]{#1}%
\providecommand \citenamefont [1]{#1}%
\providecommand \href@noop [0]{\@secondoftwo}%
\providecommand \href [0]{\begingroup \@sanitize@url \@href}%
\providecommand \@href[1]{\@@startlink{#1}\@@href}%
\providecommand \@@href[1]{\endgroup#1\@@endlink}%
\providecommand \@sanitize@url [0]{\catcode `\\12\catcode `\$12\catcode
  `\&12\catcode `\#12\catcode `\^12\catcode `\_12\catcode `\%12\relax}%
\providecommand \@@startlink[1]{}%
\providecommand \@@endlink[0]{}%
\providecommand \url  [0]{\begingroup\@sanitize@url \@url }%
\providecommand \@url [1]{\endgroup\@href {#1}{\urlprefix }}%
\providecommand \urlprefix  [0]{URL }%
\providecommand \Eprint [0]{\href }%
\providecommand \doibase [0]{http://dx.doi.org/}%
\providecommand \selectlanguage [0]{\@gobble}%
\providecommand \bibinfo  [0]{\@secondoftwo}%
\providecommand \bibfield  [0]{\@secondoftwo}%
\providecommand \translation [1]{[#1]}%
\providecommand \BibitemOpen [0]{}%
\providecommand \bibitemStop [0]{}%
\providecommand \bibitemNoStop [0]{.\EOS\space}%
\providecommand \EOS [0]{\spacefactor3000\relax}%
\providecommand \BibitemShut  [1]{\csname bibitem#1\endcsname}%
\let\auto@bib@innerbib\@empty
\bibitem [{\citenamefont {Abbott}\ \emph {et~al.}(2016)\citenamefont {Abbott}
  \emph {et~al.}}]{Abbott:2016blz}%
  \BibitemOpen
  \bibfield  {author} {\bibinfo {author} {\bibfnamefont {B.~P.}\ \bibnamefont
  {Abbott}} \emph {et~al.} (\bibinfo {collaboration} {Virgo, LIGO
  Scientific}),\ }\href {\doibase 10.1103/PhysRevLett.116.061102} {\bibfield
  {journal} {\bibinfo  {journal} {Phys. Rev. Lett.}\ }\textbf {\bibinfo
  {volume} {116}},\ \bibinfo {pages} {061102} (\bibinfo {year} {2016})},\
  \Eprint {http://arxiv.org/abs/1602.03837} {arXiv:1602.03837 [gr-qc]}
  \BibitemShut {NoStop}%
\bibitem [{\citenamefont {Abbott}\ \emph {et~al.}(2017)\citenamefont {Abbott}
  \emph {et~al.}}]{TheLIGOScientific:2017qsa}%
  \BibitemOpen
  \bibfield  {author} {\bibinfo {author} {\bibfnamefont {B.~P.}\ \bibnamefont
  {Abbott}} \emph {et~al.} (\bibinfo {collaboration} {Virgo, LIGO
  Scientific}),\ }\href {\doibase 10.1103/PhysRevLett.119.161101} {\bibfield
  {journal} {\bibinfo  {journal} {Phys. Rev. Lett.}\ }\textbf {\bibinfo
  {volume} {119}},\ \bibinfo {pages} {161101} (\bibinfo {year} {2017})},\
  \Eprint {http://arxiv.org/abs/1710.05832} {arXiv:1710.05832 [gr-qc]}
  \BibitemShut {NoStop}%
\bibitem [{\citenamefont {Krasnikov}(1987)}]{Krasnikov:1987yj}%
  \BibitemOpen
  \bibfield  {author} {\bibinfo {author} {\bibfnamefont {N.~V.}\ \bibnamefont
  {Krasnikov}},\ }\href {\doibase 10.1007/BF01017588} {\bibfield  {journal}
  {\bibinfo  {journal} {Theor. Math. Phys.}\ }\textbf {\bibinfo {volume}
  {73}},\ \bibinfo {pages} {1184} (\bibinfo {year} {1987})},\ \bibinfo {note}
  {[Teor. Mat. Fiz.73,235(1987)]}\BibitemShut {NoStop}%
\bibitem [{\citenamefont {Tomboulis}(1997)}]{Tomboulis:1997gg}%
  \BibitemOpen
  \bibfield  {author} {\bibinfo {author} {\bibfnamefont {E.~T.}\ \bibnamefont
  {Tomboulis}},\ }\href@noop {} {\  (\bibinfo {year} {1997})},\ \Eprint
  {http://arxiv.org/abs/hep-th/9702146} {arXiv:hep-th/9702146 [hep-th]}
  \BibitemShut {NoStop}%
\bibitem [{\citenamefont {Modesto}(2012)}]{Modesto:2011kw}%
  \BibitemOpen
  \bibfield  {author} {\bibinfo {author} {\bibfnamefont {L.}~\bibnamefont
  {Modesto}},\ }\href {\doibase 10.1103/PhysRevD.86.044005} {\bibfield
  {journal} {\bibinfo  {journal} {Phys. Rev.}\ }\textbf {\bibinfo {volume}
  {D86}},\ \bibinfo {pages} {044005} (\bibinfo {year} {2012})},\ \Eprint
  {http://arxiv.org/abs/1107.2403} {arXiv:1107.2403 [hep-th]} \BibitemShut
  {NoStop}%
\bibitem [{\citenamefont {Biswas}\ \emph {et~al.}(2012)\citenamefont {Biswas},
  \citenamefont {Gerwick}, \citenamefont {Koivisto},\ and\ \citenamefont
  {Mazumdar}}]{Biswas:2011ar}%
  \BibitemOpen
  \bibfield  {author} {\bibinfo {author} {\bibfnamefont {T.}~\bibnamefont
  {Biswas}}, \bibinfo {author} {\bibfnamefont {E.}~\bibnamefont {Gerwick}},
  \bibinfo {author} {\bibfnamefont {T.}~\bibnamefont {Koivisto}}, \ and\
  \bibinfo {author} {\bibfnamefont {A.}~\bibnamefont {Mazumdar}},\ }\href
  {\doibase 10.1103/PhysRevLett.108.031101} {\bibfield  {journal} {\bibinfo
  {journal} {Phys. Rev. Lett.}\ }\textbf {\bibinfo {volume} {108}},\ \bibinfo
  {pages} {031101} (\bibinfo {year} {2012})},\ \Eprint
  {http://arxiv.org/abs/1110.5249} {arXiv:1110.5249 [gr-qc]} \BibitemShut
  {NoStop}%
\bibitem [{\citenamefont {Talaganis}\ \emph {et~al.}(2015)\citenamefont
  {Talaganis}, \citenamefont {Biswas},\ and\ \citenamefont
  {Mazumdar}}]{Talaganis:2014ida}%
  \BibitemOpen
  \bibfield  {author} {\bibinfo {author} {\bibfnamefont {S.}~\bibnamefont
  {Talaganis}}, \bibinfo {author} {\bibfnamefont {T.}~\bibnamefont {Biswas}}, \
  and\ \bibinfo {author} {\bibfnamefont {A.}~\bibnamefont {Mazumdar}},\ }\href
  {\doibase 10.1088/0264-9381/32/21/215017} {\bibfield  {journal} {\bibinfo
  {journal} {Class. Quant. Grav.}\ }\textbf {\bibinfo {volume} {32}},\ \bibinfo
  {pages} {215017} (\bibinfo {year} {2015})},\ \Eprint
  {http://arxiv.org/abs/1412.3467} {arXiv:1412.3467 [hep-th]} \BibitemShut
  {NoStop}%
\bibitem [{\citenamefont {Modesto}\ and\ \citenamefont
  {Rachwal}(2015)}]{Modesto:2015lna}%
  \BibitemOpen
  \bibfield  {author} {\bibinfo {author} {\bibfnamefont {L.}~\bibnamefont
  {Modesto}}\ and\ \bibinfo {author} {\bibfnamefont {L.}~\bibnamefont
  {Rachwal}},\ }\href {\doibase 10.1016/j.nuclphysb.2015.09.006} {\bibfield
  {journal} {\bibinfo  {journal} {Nucl. Phys.}\ }\textbf {\bibinfo {volume}
  {B900}},\ \bibinfo {pages} {147} (\bibinfo {year} {2015})},\ \Eprint
  {http://arxiv.org/abs/1503.00261} {arXiv:1503.00261 [hep-th]} \BibitemShut
  {NoStop}%
\bibitem [{\citenamefont {Don\'a}\ \emph {et~al.}(2015)\citenamefont {Don\'a},
  \citenamefont {Giaccari}, \citenamefont {Modesto}, \citenamefont {Rachwal},\
  and\ \citenamefont {Zhu}}]{Dona:2015tra}%
  \BibitemOpen
  \bibfield  {author} {\bibinfo {author} {\bibfnamefont {P.}~\bibnamefont
  {Don\'a}}, \bibinfo {author} {\bibfnamefont {S.}~\bibnamefont {Giaccari}},
  \bibinfo {author} {\bibfnamefont {L.}~\bibnamefont {Modesto}}, \bibinfo
  {author} {\bibfnamefont {L.}~\bibnamefont {Rachwal}}, \ and\ \bibinfo
  {author} {\bibfnamefont {Y.}~\bibnamefont {Zhu}},\ }\href {\doibase
  10.1007/JHEP08(2015)038} {\bibfield  {journal} {\bibinfo  {journal} {JHEP}\
  }\textbf {\bibinfo {volume} {08}},\ \bibinfo {pages} {038} (\bibinfo {year}
  {2015})},\ \Eprint {http://arxiv.org/abs/1506.04589} {arXiv:1506.04589
  [hep-th]} \BibitemShut {NoStop}%
\bibitem [{\citenamefont {Talaganis}\ and\ \citenamefont
  {Mazumdar}(2016)}]{Talaganis:2016ovm}%
  \BibitemOpen
  \bibfield  {author} {\bibinfo {author} {\bibfnamefont {S.}~\bibnamefont
  {Talaganis}}\ and\ \bibinfo {author} {\bibfnamefont {A.}~\bibnamefont
  {Mazumdar}},\ }\href {\doibase 10.1088/0264-9381/33/14/145005} {\bibfield
  {journal} {\bibinfo  {journal} {Class. Quant. Grav.}\ }\textbf {\bibinfo
  {volume} {33}},\ \bibinfo {pages} {145005} (\bibinfo {year} {2016})},\
  \Eprint {http://arxiv.org/abs/1603.03440} {arXiv:1603.03440 [hep-th]}
  \BibitemShut {NoStop}%
\bibitem [{\citenamefont {Briscese}\ \emph {et~al.}(2014)\citenamefont
  {Briscese}, \citenamefont {Modesto},\ and\ \citenamefont
  {Tsujikawa}}]{Briscese:2013lna}%
  \BibitemOpen
  \bibfield  {author} {\bibinfo {author} {\bibfnamefont {F.}~\bibnamefont
  {Briscese}}, \bibinfo {author} {\bibfnamefont {L.}~\bibnamefont {Modesto}}, \
  and\ \bibinfo {author} {\bibfnamefont {S.}~\bibnamefont {Tsujikawa}},\ }\href
  {\doibase 10.1103/PhysRevD.89.024029} {\bibfield  {journal} {\bibinfo
  {journal} {Phys. Rev.}\ }\textbf {\bibinfo {volume} {D89}},\ \bibinfo {pages}
  {024029} (\bibinfo {year} {2014})},\ \Eprint {http://arxiv.org/abs/1308.1413}
  {arXiv:1308.1413 [hep-th]} \BibitemShut {NoStop}%
\bibitem [{\citenamefont {Koshelev}\ \emph {et~al.}(2016)\citenamefont
  {Koshelev}, \citenamefont {Modesto}, \citenamefont {Rachwal},\ and\
  \citenamefont {Starobinsky}}]{Koshelev:2016xqb}%
  \BibitemOpen
  \bibfield  {author} {\bibinfo {author} {\bibfnamefont {A.~S.}\ \bibnamefont
  {Koshelev}}, \bibinfo {author} {\bibfnamefont {L.}~\bibnamefont {Modesto}},
  \bibinfo {author} {\bibfnamefont {L.}~\bibnamefont {Rachwal}}, \ and\
  \bibinfo {author} {\bibfnamefont {A.~A.}\ \bibnamefont {Starobinsky}},\
  }\href {\doibase 10.1007/JHEP11(2016)067} {\bibfield  {journal} {\bibinfo
  {journal} {JHEP}\ }\textbf {\bibinfo {volume} {11}},\ \bibinfo {pages} {067}
  (\bibinfo {year} {2016})},\ \Eprint {http://arxiv.org/abs/1604.03127}
  {arXiv:1604.03127 [hep-th]} \BibitemShut {NoStop}%
\bibitem [{\citenamefont {Biswas}\ \emph {et~al.}(2006)\citenamefont {Biswas},
  \citenamefont {Mazumdar},\ and\ \citenamefont {Siegel}}]{Biswas:2005qr}%
  \BibitemOpen
  \bibfield  {author} {\bibinfo {author} {\bibfnamefont {T.}~\bibnamefont
  {Biswas}}, \bibinfo {author} {\bibfnamefont {A.}~\bibnamefont {Mazumdar}}, \
  and\ \bibinfo {author} {\bibfnamefont {W.}~\bibnamefont {Siegel}},\ }\href
  {\doibase 10.1088/1475-7516/2006/03/009} {\bibfield  {journal} {\bibinfo
  {journal} {JCAP}\ }\textbf {\bibinfo {volume} {0603}},\ \bibinfo {pages}
  {009} (\bibinfo {year} {2006})},\ \Eprint
  {http://arxiv.org/abs/hep-th/0508194} {arXiv:hep-th/0508194 [hep-th]}
  \BibitemShut {NoStop}%
\bibitem [{\citenamefont {Biswas}\ \emph {et~al.}(2010)\citenamefont {Biswas},
  \citenamefont {Koivisto},\ and\ \citenamefont {Mazumdar}}]{Biswas:2010zk}%
  \BibitemOpen
  \bibfield  {author} {\bibinfo {author} {\bibfnamefont {T.}~\bibnamefont
  {Biswas}}, \bibinfo {author} {\bibfnamefont {T.}~\bibnamefont {Koivisto}}, \
  and\ \bibinfo {author} {\bibfnamefont {A.}~\bibnamefont {Mazumdar}},\ }\href
  {\doibase 10.1088/1475-7516/2010/11/008} {\bibfield  {journal} {\bibinfo
  {journal} {JCAP}\ }\textbf {\bibinfo {volume} {1011}},\ \bibinfo {pages}
  {008} (\bibinfo {year} {2010})},\ \Eprint {http://arxiv.org/abs/1005.0590}
  {arXiv:1005.0590 [hep-th]} \BibitemShut {NoStop}%
\bibitem [{\citenamefont {Conroy}\ \emph {et~al.}(2017)\citenamefont {Conroy},
  \citenamefont {Koshelev},\ and\ \citenamefont {Mazumdar}}]{Conroy:2016sac}%
  \BibitemOpen
  \bibfield  {author} {\bibinfo {author} {\bibfnamefont {A.}~\bibnamefont
  {Conroy}}, \bibinfo {author} {\bibfnamefont {A.~S.}\ \bibnamefont
  {Koshelev}}, \ and\ \bibinfo {author} {\bibfnamefont {A.}~\bibnamefont
  {Mazumdar}},\ }\href {\doibase 10.1088/1475-7516/2017/01/017} {\bibfield
  {journal} {\bibinfo  {journal} {JCAP}\ }\textbf {\bibinfo {volume} {1701}},\
  \bibinfo {pages} {017} (\bibinfo {year} {2017})},\ \Eprint
  {http://arxiv.org/abs/1605.02080} {arXiv:1605.02080 [gr-qc]} \BibitemShut
  {NoStop}%
\bibitem [{\citenamefont {Frolov}(2015)}]{Frolov:2015bta}%
  \BibitemOpen
  \bibfield  {author} {\bibinfo {author} {\bibfnamefont {V.~P.}\ \bibnamefont
  {Frolov}},\ }\href {\doibase 10.1103/PhysRevLett.115.051102} {\bibfield
  {journal} {\bibinfo  {journal} {Phys. Rev. Lett.}\ }\textbf {\bibinfo
  {volume} {115}},\ \bibinfo {pages} {051102} (\bibinfo {year} {2015})},\
  \Eprint {http://arxiv.org/abs/1505.00492} {arXiv:1505.00492 [hep-th]}
  \BibitemShut {NoStop}%
\bibitem [{\citenamefont {Conroy}\ \emph
  {et~al.}(2015{\natexlab{a}})\citenamefont {Conroy}, \citenamefont
  {Mazumdar},\ and\ \citenamefont {Teimouri}}]{Conroy:2015wfa}%
  \BibitemOpen
  \bibfield  {author} {\bibinfo {author} {\bibfnamefont {A.}~\bibnamefont
  {Conroy}}, \bibinfo {author} {\bibfnamefont {A.}~\bibnamefont {Mazumdar}}, \
  and\ \bibinfo {author} {\bibfnamefont {A.}~\bibnamefont {Teimouri}},\ }\href
  {\doibase 10.1103/PhysRevLett.114.201101} {\bibfield  {journal} {\bibinfo
  {journal} {Phys. Rev. Lett.}\ }\textbf {\bibinfo {volume} {114}},\ \bibinfo
  {pages} {201101} (\bibinfo {year} {2015}{\natexlab{a}})},\ \Eprint
  {http://arxiv.org/abs/1503.05568} {arXiv:1503.05568 [hep-th]} \BibitemShut
  {NoStop}%
\bibitem [{\citenamefont {Feynman}(1996)}]{Feynman:1996kb}%
  \BibitemOpen
  \bibfield  {author} {\bibinfo {author} {\bibfnamefont {R.~P.}\ \bibnamefont
  {Feynman}},\ }\href@noop {} {\emph {\bibinfo {title} {{Feynman lectures on
  gravitation}}}},\ edited by\ \bibinfo {editor} {\bibfnamefont {F.~B.}\
  \bibnamefont {Morinigo}}, \bibinfo {editor} {\bibfnamefont {W.~G.}\
  \bibnamefont {Wagner}}, \ and\ \bibinfo {editor} {\bibfnamefont
  {B.}~\bibnamefont {Hatfield}}\ (\bibinfo {year} {1996})\BibitemShut {NoStop}%
\bibitem [{\citenamefont {de~Andrade}\ \emph {et~al.}(2000)\citenamefont
  {de~Andrade}, \citenamefont {Guillen},\ and\ \citenamefont
  {Pereira}}]{deAndrade:2000kr}%
  \BibitemOpen
  \bibfield  {author} {\bibinfo {author} {\bibfnamefont {V.~C.}\ \bibnamefont
  {de~Andrade}}, \bibinfo {author} {\bibfnamefont {L.~C.~T.}\ \bibnamefont
  {Guillen}}, \ and\ \bibinfo {author} {\bibfnamefont {J.~G.}\ \bibnamefont
  {Pereira}},\ }\href {\doibase 10.1103/PhysRevLett.84.4533} {\bibfield
  {journal} {\bibinfo  {journal} {Phys. Rev. Lett.}\ }\textbf {\bibinfo
  {volume} {84}},\ \bibinfo {pages} {4533} (\bibinfo {year} {2000})},\ \Eprint
  {http://arxiv.org/abs/gr-qc/0003100} {arXiv:gr-qc/0003100 [gr-qc]}
  \BibitemShut {NoStop}%
\bibitem [{\citenamefont {{Isham}}\ \emph {et~al.}(1971)\citenamefont
  {{Isham}}, \citenamefont {{Salam}},\ and\ \citenamefont
  {{Strathdee}}}]{1971AnPhy..62...98I}%
  \BibitemOpen
  \bibfield  {author} {\bibinfo {author} {\bibfnamefont {C.~J.}\ \bibnamefont
  {{Isham}}}, \bibinfo {author} {\bibfnamefont {A.}~\bibnamefont {{Salam}}}, \
  and\ \bibinfo {author} {\bibfnamefont {J.}~\bibnamefont {{Strathdee}}},\
  }\href {\doibase 10.1016/0003-4916(71)90269-7} {\bibfield  {journal}
  {\bibinfo  {journal} {Annals of Physics}\ }\textbf {\bibinfo {volume} {62}},\
  \bibinfo {pages} {98} (\bibinfo {year} {1971})}\BibitemShut {NoStop}%
\bibitem [{\citenamefont {Hehl}\ \emph {et~al.}(1995)\citenamefont {Hehl},
  \citenamefont {McCrea}, \citenamefont {Mielke},\ and\ \citenamefont
  {Ne'eman}}]{Hehl:1994ue}%
  \BibitemOpen
  \bibfield  {author} {\bibinfo {author} {\bibfnamefont {F.~W.}\ \bibnamefont
  {Hehl}}, \bibinfo {author} {\bibfnamefont {J.~D.}\ \bibnamefont {McCrea}},
  \bibinfo {author} {\bibfnamefont {E.~W.}\ \bibnamefont {Mielke}}, \ and\
  \bibinfo {author} {\bibfnamefont {Y.}~\bibnamefont {Ne'eman}},\ }\href
  {\doibase 10.1016/0370-1573(94)00111-F} {\bibfield  {journal} {\bibinfo
  {journal} {Phys. Rept.}\ }\textbf {\bibinfo {volume} {258}},\ \bibinfo
  {pages} {1} (\bibinfo {year} {1995})},\ \Eprint
  {http://arxiv.org/abs/gr-qc/9402012} {arXiv:gr-qc/9402012 [gr-qc]}
  \BibitemShut {NoStop}%
\bibitem [{\citenamefont {Beltran~Jimenez}\ \emph {et~al.}(2017)\citenamefont
  {Beltran~Jimenez}, \citenamefont {Heisenberg},\ and\ \citenamefont
  {Koivisto}}]{us}%
  \BibitemOpen
  \bibfield  {author} {\bibinfo {author} {\bibfnamefont {J.}~\bibnamefont
  {Beltran~Jimenez}}, \bibinfo {author} {\bibfnamefont {L.}~\bibnamefont
  {Heisenberg}}, \ and\ \bibinfo {author} {\bibfnamefont {T.}~\bibnamefont
  {Koivisto}},\ }\href@noop {} {\  (\bibinfo {year} {2017})},\ \Eprint
  {http://arxiv.org/abs/1710.03116} {arXiv:1710.03116 [gr-qc]} \BibitemShut
  {NoStop}%
\bibitem [{\citenamefont {Nester}\ and\ \citenamefont
  {Yo}(1999)}]{Nester:1998mp}%
  \BibitemOpen
  \bibfield  {author} {\bibinfo {author} {\bibfnamefont {J.~M.}\ \bibnamefont
  {Nester}}\ and\ \bibinfo {author} {\bibfnamefont {H.-J.}\ \bibnamefont
  {Yo}},\ }\href@noop {} {\bibfield  {journal} {\bibinfo  {journal} {Chin. J.
  Phys.}\ }\textbf {\bibinfo {volume} {37}},\ \bibinfo {pages} {113} (\bibinfo
  {year} {1999})},\ \Eprint {http://arxiv.org/abs/gr-qc/9809049}
  {arXiv:gr-qc/9809049 [gr-qc]} \BibitemShut {NoStop}%
\bibitem [{\citenamefont {Adak}\ \emph {et~al.}(2006)\citenamefont {Adak},
  \citenamefont {Kalay},\ and\ \citenamefont {Sert}}]{Adak:2005cd}%
  \BibitemOpen
  \bibfield  {author} {\bibinfo {author} {\bibfnamefont {M.}~\bibnamefont
  {Adak}}, \bibinfo {author} {\bibfnamefont {M.}~\bibnamefont {Kalay}}, \ and\
  \bibinfo {author} {\bibfnamefont {O.}~\bibnamefont {Sert}},\ }\href {\doibase
  10.1142/S0218271806008474} {\bibfield  {journal} {\bibinfo  {journal} {Int.
  J. Mod. Phys.}\ }\textbf {\bibinfo {volume} {D15}},\ \bibinfo {pages} {619}
  (\bibinfo {year} {2006})},\ \Eprint {http://arxiv.org/abs/gr-qc/0505025}
  {arXiv:gr-qc/0505025 [gr-qc]} \BibitemShut {NoStop}%
\bibitem [{\citenamefont {Adak}\ \emph {et~al.}(2013)\citenamefont {Adak},
  \citenamefont {Sert}, \citenamefont {Kalay},\ and\ \citenamefont
  {Sari}}]{Adak:2008gd}%
  \BibitemOpen
  \bibfield  {author} {\bibinfo {author} {\bibfnamefont {M.}~\bibnamefont
  {Adak}}, \bibinfo {author} {\bibfnamefont {z.}~\bibnamefont {Sert}}, \bibinfo
  {author} {\bibfnamefont {M.}~\bibnamefont {Kalay}}, \ and\ \bibinfo {author}
  {\bibfnamefont {M.}~\bibnamefont {Sari}},\ }\href {\doibase
  10.1142/S0217751X13501674} {\bibfield  {journal} {\bibinfo  {journal} {Int.
  J. Mod. Phys.}\ }\textbf {\bibinfo {volume} {A28}},\ \bibinfo {pages}
  {1350167} (\bibinfo {year} {2013})},\ \Eprint
  {http://arxiv.org/abs/0810.2388} {arXiv:0810.2388 [gr-qc]} \BibitemShut
  {NoStop}%
\bibitem [{\citenamefont {Mol}(2017)}]{Mol:2014ooa}%
  \BibitemOpen
  \bibfield  {author} {\bibinfo {author} {\bibfnamefont {I.}~\bibnamefont
  {Mol}},\ }\href {\doibase 10.1007/s00006-016-0749-8} {\bibfield  {journal}
  {\bibinfo  {journal} {Adv. Appl. Clifford Algebras}\ }\textbf {\bibinfo
  {volume} {27}},\ \bibinfo {pages} {2607} (\bibinfo {year} {2017})},\ \Eprint
  {http://arxiv.org/abs/1406.0737} {arXiv:1406.0737 [physics.gen-ph]}
  \BibitemShut {NoStop}%
\bibitem [{\citenamefont {Kibble}(1961)}]{Kibble:1961ba}%
  \BibitemOpen
  \bibfield  {author} {\bibinfo {author} {\bibfnamefont {T.~W.~B.}\
  \bibnamefont {Kibble}},\ }\href {\doibase 10.1063/1.1703702} {\bibfield
  {journal} {\bibinfo  {journal} {J. Math. Phys.}\ }\textbf {\bibinfo {volume}
  {2}},\ \bibinfo {pages} {212} (\bibinfo {year} {1961})}\BibitemShut {NoStop}%
\bibitem [{\citenamefont {Tresguerres}\ and\ \citenamefont
  {Mielke}(2000)}]{Tresguerres:2000qn}%
  \BibitemOpen
  \bibfield  {author} {\bibinfo {author} {\bibfnamefont {R.}~\bibnamefont
  {Tresguerres}}\ and\ \bibinfo {author} {\bibfnamefont {E.~W.}\ \bibnamefont
  {Mielke}},\ }\href {\doibase 10.1103/PhysRevD.62.044004} {\bibfield
  {journal} {\bibinfo  {journal} {Phys. Rev.}\ }\textbf {\bibinfo {volume}
  {D62}},\ \bibinfo {pages} {044004} (\bibinfo {year} {2000})},\ \Eprint
  {http://arxiv.org/abs/gr-qc/0007072} {arXiv:gr-qc/0007072 [gr-qc]}
  \BibitemShut {NoStop}%
\bibitem [{\citenamefont {Golovnev}\ \emph {et~al.}(2017)\citenamefont
  {Golovnev}, \citenamefont {Koivisto},\ and\ \citenamefont
  {Sandstad}}]{Golovnev:2017dox}%
  \BibitemOpen
  \bibfield  {author} {\bibinfo {author} {\bibfnamefont {A.}~\bibnamefont
  {Golovnev}}, \bibinfo {author} {\bibfnamefont {T.}~\bibnamefont {Koivisto}},
  \ and\ \bibinfo {author} {\bibfnamefont {M.}~\bibnamefont {Sandstad}},\
  }\href {\doibase 10.1088/1361-6382/aa7830} {\bibfield  {journal} {\bibinfo
  {journal} {Class. Quant. Grav.}\ }\textbf {\bibinfo {volume} {34}},\ \bibinfo
  {pages} {145013} (\bibinfo {year} {2017})},\ \Eprint
  {http://arxiv.org/abs/1701.06271} {arXiv:1701.06271 [gr-qc]} \BibitemShut
  {NoStop}%
\bibitem [{\citenamefont {Koivisto}(2018)}]{Koivisto:2018aip}%
  \BibitemOpen
  \bibfield  {author} {\bibinfo {author} {\bibfnamefont {T.}~\bibnamefont
  {Koivisto}}\ }(\bibinfo {year} {2018})\ \Eprint
  {http://arxiv.org/abs/1802.00650} {arXiv:1802.00650 [gr-qc]} \BibitemShut
  {NoStop}%
\bibitem [{\citenamefont {Van~Nieuwenhuizen}(1973)}]{VanNieuwenhuizen:1973fi}%
  \BibitemOpen
  \bibfield  {author} {\bibinfo {author} {\bibfnamefont {P.}~\bibnamefont
  {Van~Nieuwenhuizen}},\ }\href {\doibase 10.1016/0550-3213(73)90194-6}
  {\bibfield  {journal} {\bibinfo  {journal} {Nucl. Phys.}\ }\textbf {\bibinfo
  {volume} {B60}},\ \bibinfo {pages} {478} (\bibinfo {year}
  {1973})}\BibitemShut {NoStop}%
\bibitem [{\citenamefont {Alvarez}\ \emph {et~al.}(2006)\citenamefont
  {Alvarez}, \citenamefont {Blas}, \citenamefont {Garriga},\ and\ \citenamefont
  {Verdaguer}}]{Alvarez:2006uu}%
  \BibitemOpen
  \bibfield  {author} {\bibinfo {author} {\bibfnamefont {E.}~\bibnamefont
  {Alvarez}}, \bibinfo {author} {\bibfnamefont {D.}~\bibnamefont {Blas}},
  \bibinfo {author} {\bibfnamefont {J.}~\bibnamefont {Garriga}}, \ and\
  \bibinfo {author} {\bibfnamefont {E.}~\bibnamefont {Verdaguer}},\ }\href
  {\doibase 10.1016/j.nuclphysb.2006.08.003} {\bibfield  {journal} {\bibinfo
  {journal} {Nucl. Phys.}\ }\textbf {\bibinfo {volume} {B756}},\ \bibinfo
  {pages} {148} (\bibinfo {year} {2006})},\ \Eprint
  {http://arxiv.org/abs/hep-th/0606019} {arXiv:hep-th/0606019 [hep-th]}
  \BibitemShut {NoStop}%
\bibitem [{\citenamefont {Conroy}\ \emph
  {et~al.}(2015{\natexlab{b}})\citenamefont {Conroy}, \citenamefont {Koivisto},
  \citenamefont {Mazumdar},\ and\ \citenamefont {Teimouri}}]{Conroy:2014eja}%
  \BibitemOpen
  \bibfield  {author} {\bibinfo {author} {\bibfnamefont {A.}~\bibnamefont
  {Conroy}}, \bibinfo {author} {\bibfnamefont {T.}~\bibnamefont {Koivisto}},
  \bibinfo {author} {\bibfnamefont {A.}~\bibnamefont {Mazumdar}}, \ and\
  \bibinfo {author} {\bibfnamefont {A.}~\bibnamefont {Teimouri}},\ }\href
  {\doibase 10.1088/0264-9381/32/1/015024} {\bibfield  {journal} {\bibinfo
  {journal} {Class. Quant. Grav.}\ }\textbf {\bibinfo {volume} {32}},\ \bibinfo
  {pages} {015024} (\bibinfo {year} {2015}{\natexlab{b}})},\ \Eprint
  {http://arxiv.org/abs/1406.4998} {arXiv:1406.4998 [hep-th]} \BibitemShut
  {NoStop}%
\bibitem [{\citenamefont {Chen}\ \emph {et~al.}(1998)\citenamefont {Chen},
  \citenamefont {Nester},\ and\ \citenamefont {Yo}}]{Chen:1998ad}%
  \BibitemOpen
  \bibfield  {author} {\bibinfo {author} {\bibfnamefont {H.}~\bibnamefont
  {Chen}}, \bibinfo {author} {\bibfnamefont {J.~M.}\ \bibnamefont {Nester}}, \
  and\ \bibinfo {author} {\bibfnamefont {H.-J.}\ \bibnamefont {Yo}},\
  }\bibfield  {booktitle} {\emph {\bibinfo {booktitle} {{Gauge theories of
  gravitation. Proceedings, Workshop, Jadwisin, Poland, September 4-10,
  1997}}},\ }\href@noop {} {\bibfield  {journal} {\bibinfo  {journal} {Acta
  Phys. Polon.}\ }\textbf {\bibinfo {volume} {B29}},\ \bibinfo {pages} {961}
  (\bibinfo {year} {1998})}\BibitemShut {NoStop}%
\bibitem [{\citenamefont {Deser}\ \emph {et~al.}(2015)\citenamefont {Deser},
  \citenamefont {Izumi}, \citenamefont {Ong},\ and\ \citenamefont
  {Waldron}}]{Deser:2014fta}%
  \BibitemOpen
  \bibfield  {author} {\bibinfo {author} {\bibfnamefont {S.}~\bibnamefont
  {Deser}}, \bibinfo {author} {\bibfnamefont {K.}~\bibnamefont {Izumi}},
  \bibinfo {author} {\bibfnamefont {Y.~C.}\ \bibnamefont {Ong}}, \ and\
  \bibinfo {author} {\bibfnamefont {A.}~\bibnamefont {Waldron}},\ }\href
  {\doibase 10.1142/S0217732315400064} {\bibfield  {journal} {\bibinfo
  {journal} {Mod. Phys. Lett.}\ }\textbf {\bibinfo {volume} {A30}},\ \bibinfo
  {pages} {1540006} (\bibinfo {year} {2015})},\ \Eprint
  {http://arxiv.org/abs/1410.2289} {arXiv:1410.2289 [hep-th]} \BibitemShut
  {NoStop}%
\bibitem [{\citenamefont {Deser}\ and\ \citenamefont
  {Woodard}(2007)}]{Deser:2007jk}%
  \BibitemOpen
  \bibfield  {author} {\bibinfo {author} {\bibfnamefont {S.}~\bibnamefont
  {Deser}}\ and\ \bibinfo {author} {\bibfnamefont {R.~P.}\ \bibnamefont
  {Woodard}},\ }\href {\doibase 10.1103/PhysRevLett.99.111301} {\bibfield
  {journal} {\bibinfo  {journal} {Phys. Rev. Lett.}\ }\textbf {\bibinfo
  {volume} {99}},\ \bibinfo {pages} {111301} (\bibinfo {year} {2007})},\
  \Eprint {http://arxiv.org/abs/0706.2151} {arXiv:0706.2151 [astro-ph]}
  \BibitemShut {NoStop}%
\bibitem [{\citenamefont {Lammerzahl}(1996)}]{Lammerzahl:1996se}%
  \BibitemOpen
  \bibfield  {author} {\bibinfo {author} {\bibfnamefont {C.}~\bibnamefont
  {Lammerzahl}},\ }\href {\doibase 10.1007/BF02113157} {\bibfield  {journal}
  {\bibinfo  {journal} {Gen. Rel. Grav.}\ }\textbf {\bibinfo {volume} {28}},\
  \bibinfo {pages} {1043} (\bibinfo {year} {1996})},\ \Eprint
  {http://arxiv.org/abs/gr-qc/9605065} {arXiv:gr-qc/9605065 [gr-qc]}
  \BibitemShut {NoStop}%
\bibitem [{\citenamefont {Chiao}(2003)}]{Chiao:2003es}%
  \BibitemOpen
  \bibfield  {author} {\bibinfo {author} {\bibfnamefont {R.~Y.}\ \bibnamefont
  {Chiao}},\ }\href@noop {} {\  (\bibinfo {year} {2003})},\ \Eprint
  {http://arxiv.org/abs/gr-qc/0303100} {arXiv:gr-qc/0303100 [gr-qc]}
  \BibitemShut {NoStop}%
\end{thebibliography}%
	
\end{document}